**Choosing the Causal Estimand for Propensity Score Analysis of Observational Studies**

Noah Greifer
Institute for Quantitative Social Science, Harvard University
ngreifer@iq.harvard.edu

Elizabeth A. Stuart
Department of Mental Health, Johns Hopkins Bloomberg School of Public Health; Department of Health Policy and Management, Johns Hopkins Bloomberg School of Public Health
estuart@jhu.edu

ABSTRACT

Matching and weighting methods for observational studies involve the choice of an estimand, the causal effect with reference to a specific target population. Commonly used estimands include the average treatment effect in the treated (ATT), the average treatment effect in the untreated (ATU), the average treatment effect in the population (ATE), and the average treatment effect in the overlap (i.e., equipoise population; ATO). Each estimand has its own assumptions, interpretation, and statistical methods that can be used to estimate it. This article provides guidance on selecting and interpreting an estimand to help medical researchers correctly implement statistical methods used to estimate causal effects in observational studies and to help audiences correctly interpret the results and limitations of these studies. The interpretations of the estimands resulting from regression and instrumental variable analyses are also discussed. Choosing an estimand carefully is essential for making valid inferences from the analysis of observational data and ensuring results are replicable and useful for practitioners.

BACKGROUND

Medical researchers are often interested in the effect of a treatment, such as a new drug or method of surgery, on patients' health outcomes. When patients are not randomly assigned to treatment, such as when analyzing secondary data in the form of healthcare claims databases or patient medical records, differences between treated and untreated patients on characteristics prognostic of clinical endpoints will yield confounding bias in a simple unadjusted comparison of the average observed outcomes for the treated and untreated groups. Propensity score-based methods, such as propensity score matching (PSM) (1) and inverse probability weighting (IPW) (2), are popular methods to address this bias due to confounding by equating the observed patient characteristics across the treated and untreated units (3).

When estimating treatment effects, it is important to consider the population for which the effect is being estimated. The "estimand" refers to the effect of interest with consideration of a particular target population or subpopulation. The average treatment effect in the treated (ATT), for example, is the average treatment effect among units like those who actually received treatment in the study population – i.e., the difference between the average outcomes observed for the treated patients and the average outcomes they would have experienced had they instead not been treated. The average treatment effect in the untreated (ATU) is the average treatment effect among units like those who did not receive treatment in the study population. The average treatment effect in the population (ATE) is the average treatment effect among all eligible patients in the study population – i.e., the difference in average outcomes were they all to be treated vs. were they all not to be treated.

Consideration of the estimand is important not just for interpreting the results of an analysis but also for choosing the specific statistical method used to estimate the effect (4). Some methods, such as PSM in its most commonly used form, cannot target the ATE, and so are inappropriate when the ATE is of interest. Despite this, PSM is a widely used method in the medical literature, and studies often compare the results from matching with the results from other methods that target a different estimand, such as outcome regression or IPW (5,6).

This article provides a guide for researchers to consider how to choose an estimand to help answer their specific questions of interest and how that choice should motivate the methods used. This article also aims to help consumers of medical research interpret the results of statistical analyses of observational studies, in particular, to clarify for what population a study is estimating effects and to identify the extent to which effects estimated using different methods can be compared.

## Terminology and Notation

In this article, we use the terms "treated" and "untreated" to refer to the two groups under study, though the messages in this paper apply to scenarios beyond simple treated vs. untreated comparisons and can be used when comparing any two groups, policies, exposures, or medical practices. For example, researchers may want to compare a new drug to a placebo, a new drug to an existing drug, a new procedure to a standard of care, or exposure at some level of a pollutant to exposure at another level. Extensions exist for multi-category and continuously-valued treatments as well, which we discuss in the appendix. We consider the "treated" to be an active group, such as a new drug or experimental surgical practice, and the "untreated" to be an inactive group, such as a placebo or standard care practice.

We use potential outcomes notation to define causal effects. The potential outcomes are denoted as $Y^0$ and $Y^1$ for the potential outcomes in the absence and presence of treatment, respectively. For example, for patient $i$ in a study of the effect of a new drug vs. the standard of care on blood pressure, $Y_i^0$ would be the patient's blood pressure were they to receive the standard of care, and $Y_i^1$ their blood pressure were they to take the new drug. The potential outcomes for an individual can be thought of as existing prior to treatment receipt; the treatment received *reveals* one of the potential outcomes, the one corresponding to the treatment received.

The individual causal effect (ICE) for a single patient $i$ is $Y_i^1 - Y_i^0$, the difference between the outcome were the patient to be treated and the outcome were the patient not to be treated. The ICE is the "holy grail" of causal inference research and precision medicine because it would allow a doctor to identify, for each patient seen, exactly which treatment would yield the best outcome. In practice, though, only one potential outcome for each patient is observed: the one corresponding to the treatment actually received (7). When certain assumptions are met (which we discuss briefly below), one can estimate an *average* treatment effect, which is the average of the ICEs for some population of patients.

Consider the example population presented in Table 1 for a hypothetical non-experimental study of the effect of a new form of surgery ($T = 1$) vs. the standard surgical practice ($T = 0$) on a quality-of-life scale ranging from 0 to 100, where each row corresponds to a patient, $Y^1$ and $Y^0$ are the potential outcomes under the new and standard surgical practices, respectively, and $X$ is a binary risk factor that affects both the assignment to treatment and the outcome. The italicized potential outcomes are those not observed by the researcher because they correspond to the treatment not received. The ICEs are the causal effects for each patient, also unobserved by the researcher. Note that the ICEs tend to be more positive for patients with the risk factor absent ($X = 0$), and these patients are also more likely to be treated.

The average of the ICEs across all members of the population, which is called the ATE and represented symbolically as $E[Y^1 - Y^0]$, is 5 in this example, a modest improvement in quality of life if all patients were to receive the new procedure vs. were they to receive the standard surgical procedure. For the treated patients ($T = 1$), those who actually received the procedure, the average of the ICEs is 20, a more significant improvement; this quantity is the ATT, represented symbolically as $E[Y^1 - Y^0|T = 1]$. For the untreated patients ($T = 0$), however, the average of the ICEs is -5, i.e., the new surgical procedure would have yielded worse outcomes than the standard procedure they received. This quantity is the ATU, represented symbolically as $E[Y^1 - Y^0|T = 0]$. These quantities differ because patient ICEs depend on patient characteristics (in this case, $X$), which vary across treated and untreated patients differentially when patients are not randomly assigned to treatment.

It is critical to articulate which estimand is the target of a study because they can differ both in value and in interpretation, and certain statistical methods estimate some estimands and not others. In this article, we provide a rationale for the choice of these and other estimands and characterize the statistical methods used to estimate each of them.

A note on other effect measures

Although our discussion of causal effects focuses on the difference in means for a continuous outcome, other effect measures are possible with binary and survival outcomes; for example, the ATE for a binary outcome on the relative risk scale would be $P(Y^1 = 1)/P(Y^0 = 1)$, and the ATT on the relative risk/risk ratio scale would be $P(Y^1 = 1|T = 1)/P(Y^0 = 1|T = 1)$. Corresponding estimands for the odds ratio and hazard ratio can be defined similarly. Rather than understanding these contrasts as averages of unit-specific contrast, they should be understood as contrasts between averaged potential outcomes, where the average is taken over a specified patient population. The concepts described herein would still be relevant on these scales.

Table 1. A population with confounder $X$, treatment $T$, and potential outcomes $Y^1$ and $Y^0$.

| $X$ | $T$ | $Y^1$ | $Y^0$ | ICE ($Y^1 - Y^0$) | | |
|---|---|---|---|---|---|---|
| 0 | 1 | 80 | *60* | *20* | | |
| 0 | 1 | 80 | *70* | *10* | ATT = 20 | |
| 0 | 1 | 60 | *10* | *50* | | |
| 1 | 1 | 30 | *30* | *0* | | |
| 0 | 0 | *50* | 40 | *10* | | ATE = 5 |
| 0 | 0 | *30* | 40 | *-10* | | |
| 1 | 0 | *70* | 70 | *0* | ATU = -5 | |
| 1 | 0 | *60* | 50 | *10* | | |
| 1 | 0 | *50* | 80 | *-30* | | |
| 1 | 0 | *50* | 60 | *-10* | | |

Notes: ICE - individual causal effect; ATT - average treatment effect in the treated; ATU - average treatment effect in the untreated; ATE - average treatment effect in the population

Assumptions for Causal Inference

There are several causal assumptions required to be able to estimate average treatment effects from observational data, which we briefly summarize here; more detail can be found in the Appendix. These assumptions must be met to interpret an effect as causal when using matching, weighting, or regression to adjust for confounding by measured confounders. The first assumption is no unmeasured confounding, or "conditional exchangeability," which requires that all confounders of the treatment-outcome relationship, i.e., all common causes of treatment assignment and the outcome, are used in the analysis (8). The second is positivity, which requires that all patients have a nonzero probability of receiving either treatment (1,8). The third is the stable unit treatment value assumption (SUTVA), which requires that there are no unmeasured versions of treatment and that the treatment status of any other patient does not affect a patient's outcomes (9). The strictness of each assumption differs depending on the estimand targeted, as described in the Appendix.

Propensity Score Analysis

When the above assumptions are met, it is possible for a causal effect to be estimated using propensity score methods, include matching and weighting. Propensity score methods aim to reduce bias due to imbalances in confounding variables by equating the distributions of covariates in the treatment groups under study (3). Matching methods do so by dropping units from the sample or partitioning units into pairs or subclasses, and weighting methods do so by weighting the units so that the weighted distributions are similar between treatment groups (10,11). These methods often incorporate the propensity score $\pi$, a one-dimensional summary of the variables to be adjusted for, computed as the probability of receiving treatment given those variables, i.e., $\pi_i = P(T = 1|X_i)$ for a unit $i$. The most common matching method is 1:1 propensity score matching, in which pairs of units are found with similar values of the propensity score, and unpaired units are dropped from the sample (12). The most common weighting method is inverse probability weighting, which computes weights as the inverse of the probability of receiving the treatment actually received (13). As we discuss later, though, these methods do not necessarily target the same estimand.

Propensity score analysis works by first estimating the propensity score (typically using logistic regression or machine-learning based methods), performing the matching or weighting using the propensity score, and assessing whether the procedure succeeded in equating, or "balancing", the covariate distributions between the treatment groups. If not, the procedure is performed again, either by trying a different propensity score model or method of conditioning on the propensity score (e.g., matching with replacement instead of without replacement); this is done until the resulting propensity score-adjusted sample is adequate with respect to balance and the information remaining in the sample (i.e., the number of units remaining after matching or the "effective" sample size after weighting) (14).

Once an adequate adjustment specification is found, the treatment effect is estimated in the resulting sample (incorporating the weights resulting from the matching or weighting). This treatment effect generalizes to a population that resembles the resulting sample on its covariates. Different ways of specifying the matching or weighting procedure change the distribution of covariates in the resulting sample, so it is critical to understand how matching and weighting adjust the sample to ensure the interpretation of the treatment effect is consistent with the estimand actually targeted. In the following section, we describe the ATE, ATT, ATU, and other estimands and provide intuition and examples for how one should choose among them. In a later section, we describe which specifications of matching and weighting methods allow one to target each estimand. These choices are summarized in Table 2.

## A note on outcome regression

Outcome regression is an alternative to propensity score methods for adjusting for confounding and involves fitting a model (e.g., a linear, logistic, or Poisson regression model) for the outcome as a function of the treatment, patient characteristics, and possibly the propensity score. The treatment effect is often estimated as the coefficient on the treatment variable in the model. Outcome regression typically targets neither the ATE, ATT, nor ATU, but rather an unspecified estimand that is laden with assumptions and may be challenging to interpret, though there are ways to use outcome regression to target specific estimands. We discuss the limitations of outcome regression and techniques to overcome them in the Appendix.

## CHOOSING AN ESTIMAND

In a randomized trial, the treated and untreated groups will, on average, have the same distributions of patient characteristics, so the ATT, ATU, and ATE will be the same. In the absence of randomization, however, the treatment groups can have quite different distributions of characteristics, and these estimands will differ when these characteristics also relate to the treatment effect (24). Researchers then face a choice when using observational data: for whom should the treatment effect be estimated? Below, we discuss the motivations for choosing among the ATT, ATU, ATE, and average treatment effect in the "overlap" (i.e., equipoise) population (ATO) and the implications of this choice.

### Average Treatment Effect in the Treated (ATT)

The ATT is the effect of the treatment for a population of patients like those who actually received the treatment in the study. These patients may have received the treatment because treatment was decided for them based on intake characteristics such as age, body mass index, or clinical risk assessed by the provider. The question the ATT seeks to answer is "How would treated patients' outcomes differ, on average, had they, counter to fact, not received treatment?" In this sense, one can think of the ATT as the effect of *withholding* treatment from those who would otherwise receive it.

Table 2. Summary of estimands and methods for estimating them.

| Estimand | Target population | Example research question | Matching methods | Weighting methods |
|---|---|---|---|---|
| ATT | Treated patients | Should medical providers withhold treatment from those currently receiving it? | Pair matching (e.g., nearest neighbor, optimal) without a caliper (10)<br>Full matching (15)<br>Fine stratification (16) | Standardized mortality ratio weights (2) |
| ATU | Untreated (control) patients | Should medical providers extend treatment to those not currently receiving it? | Same as ATT | Same as ATT |
| ATE | Full sample/ population | Should a specific policy be applied to all eligible patients? | Full matching (15)<br>Fine stratification (16) | Inverse probability weights (13,17) |
| ATO | Clinical equipoise | Should those at clinical equipoise receive treatment?<br>Is there an effect of the treatment for some patients? | Caliper matching (10,12)<br>Coarsened exact matching (18,19)<br>Cardinality matching (20) | Overlap weights (21)<br>Matching weights (22)<br>Weight trimming (23) |

Notes: ATT - average treatment effect in the treated; ATU - average treatment effect in the untreated; ATE - average treatment effect in the population; ATO - average treatment effect in the overlap

The ATT is relevant when examining the effect of an intervention that would only be given to patients like those currently receiving it. For example, it is relevant when deciding whether a medical practice currently implemented for a group of patients should continue to be implemented for that group. It is also relevant when considering the effect of a harmful exposure or behavior, such as asbestos exposure or smoking, on patients currently exposed; the ATT then corresponds to the effect of *preventing* those patients from being exposed.

A potential challenge with the ATT is that it depends on the treatment or exposure assignment process in the population from which the data are collected. For example, if two hospitals had different prescribing practices, the treated patients in one hospital might differ from the treated patients in the other, even if the hospitals saw similar types of patients, thereby yielding two different ATTs. For this reason, it is important to document the characteristics of the treated patients in the study.

Consider the following example to motivate choosing the ATT as the target estimand. A researcher notices that doctors have been prescribing drug A for an off-label use for patients presumed to be at high risk of developing a complication from the standard care plan for managing their illness. She wants to know whether the use of drug A is actually helpful for these patients, or whether these patients would be better off with the standard care plan. Because she is interested in the effect of withholding drug A from those currently receiving it, she is interested in the ATT of drug A vs. the standard care plan. Importantly, this analysis does not provide evidence for the use of drug A for those not currently receiving it, as these patients may not resemble the treated patients and therefore may respond differently to the drug.

Average Treatment Effect in the Untreated (ATU)

The ATU is the effect of the treatment for a population like those who did not receive the treatment. The question the ATU seeks to answer is "How would untreated patients' outcomes differ, on average, had they, counter to fact, actually received treatment?" In this sense, one can think of the ATU as the effect of *expanding* treatment to patients who would otherwise not receive it.

The ATU is relevant when deciding whether a potential medical practice not currently implemented for some patients should continue not to be implemented for those patients. The ATU is most useful when considering the effect of expanding the implementation of a medical practice known to be effective for some patients to a group of patients not yet receiving it. The same concerns with respect to the treatment assignment process in different populations apply equally to the ATU; the ATU will be different for two hospitals with different treatment assignment procedures even if they see similar types of patients.

Consider the following example to motivate choosing the ATU as the target estimand. It is well known that drug B, a medication commonly prescribed for patients with a high risk of complications, is effective at lowering their risk. However, it is currently thought that the decrease in risk for low-risk patients is not substantial enough for providers to recommend drug B to them. A researcher wants to investigate whether continuing to withhold drug B from low-risk patients is preferred, or if the patients would benefit from expanding prescription of drug B to them. Because she is interested in the effect of expanding the prescription of drug B to those not currently receiving it, she is interested in the ATU of drug B vs. the standard care plan.

Average Treatment Effect in the Population (ATE)

The ATE is the effect of the treatment for the entire study population, whether its members would currently receive treatment or not. The ATE is useful for policies wide in scope or for performing cost-benefit analysis for population-wide policy. The question the ATE seeks to answer is "How would the outcomes differ, on average, were treatment given to *all* patients vs. were treatment withheld from *all* patients?" One can think of the ATE as the effect of a policy of requiring one treatment vs. requiring another, ignoring who would typically receive treatment in the absence of such a policy.

The ATE can be valuable when considering a system-wide policy for regulating a previously unregulated practice or when deciding between two care options to be implemented unilaterally. It can also be useful for assessing recommendations for when treatment decisions are not well informed, e.g., because the costs of assessing which treatment should be chosen for a given patient are high relative to the costs of the incorrect choice. Unlike the ATT and ATU, the ATE does not depend on current treatment assignment practices. The ATE involves comparing the outcomes of two (potentially unrealistic) hypothetical worlds: one in which the treatment is implemented unilaterally and one in which the treatment is withheld unilaterally. Thus, the ATE is typically not the best estimand to target when patients' benefits depend on clinical judgment but may be useful for assessing broad policies that affect all patients. The ATE may be more useful for patient-level decision-making in the context of subgroup analysis, i.e., when the patient population is narrowly defined and based on characteristics providers would use in making recommendations; we address this briefly in the Discussion section.

Consider the following example to motivate targeting the ATE. A researcher is interested in comparing the risk of complications between two drugs, drug C (a name-brand drug) and drug D (a generic drug), that aim to affect the same outcome among patients with a specific disease. She wants to know if a policy of always using one would be better than a policy of always using the other. Because she is interested in the effect of a policy affecting all eligible patients with no concern for current prescription practices, she is interested in the ATE of drug C vs. drug D.

Average Treatment Effect in the Overlap (ATO)

The ATO is the effect of the treatment in an equipoise population, i.e., a population of eligible patients for whom either treatment is currently equally implemented or there is no strong preference for one over the other (25). The patients may be those near certain clinical cut points or for whom it is ambiguous whether the benefits would outweigh the costs. Under the principle of equipoise, these are the types of patients that would be most likely to be enrolled in a clinical trial (26). The ATO answers the question, "How would the outcomes differ, on average, were patients at clinical equipoise to be given treatment vs. were treatment to be withheld?"

The ATO considers patients in a state of ambiguity rather than those for whom treatment decisions are more certain. For example, if patients with a low risk of complications are very often treated and those with a high risk are very rarely treated, the ATO would consider patients with a moderate risk, who are sometimes treated and other times not, perhaps by chance or slight preferences by providers. The patients to whom the ATO applies will be very similar to each other, but perhaps less similar to patients for whom treatment decisions are more certain. The ATO is therefore useful when clinical procedures are well understood for some patients and there is little interest in the effect of expanding or withholding treatment from them, but there are certain patients for whom it is not clear whether they should be treated or not. Unlike the other estimands, the target population of the ATO is not well defined prior to an analysis because there are many ways of defining an equipoise population, e.g., based on the degree of uncertainty with regards to treatment decisions. The target population is instead often decided by the statistical method used to estimate the treatment effect, which we will discuss in more detail in the following section. As above, after estimating the ATO, it is important to characterize the target population based on the distribution of patient characteristics so that audiences can have some idea about to whom the estimated effect applies.

Consider the following example to motivate choosing the ATO as the target estimand. Based on current practice, procedure 1 is typically indicated for patients with a low risk of complications and procedure 2 for patients with a high risk of complications. However, for those with a moderate or unknown risk, doctors tend to decide on the procedure used based on personal preference or availability. A researcher wants to know whether those patients for whom indication is ambiguous would benefit more on average from procedure 1 or procedure 2 so that doctors could make choices in a more principled way. Because she is interested in the effect of treatment for the equipoise population, she is interested in the ATO of procedure 1 vs. procedure 2.

## STATISTICAL METHODS

Different statistical methods target different estimands; if the effect of the treatment depends on any patient characteristics that affect treatment assignment, then methods targeting different estimands will be expected to yield different results, which might appear contradictory to audiences ignoring the target population of the estimand (4,27). Ideally, the choice of estimand should come first in an analysis and be based on the substantive question the researcher wishes to answer, which will imply the use of specific statistical methods to target that estimand.

Below, we describe which statistical methods commonly used in observational studies correspond to each estimand. We focus primarily on design-based methods, i.e., matching and weighting methods, which are popular in medical studies using observational data and which highlight the distinctions among the estimands, though we also briefly discuss outcome regression and instrumental variable methods. All matching and weighting methods for each estimand function in a similar way: they identify a target population and adjust each treatment group in the sample (either by weighting or selecting a subset) to resemble that target population. These methods and the estimand they target are summarized in Table 2.

### Estimating the ATT

Methods that estimate the ATT do so by using information from the untreated group to simulate what would have happened had the treated units been untreated (i.e., the potential outcomes under non-treatment for the treated). These methods, such as pair matching and standardized mortality ratio (SMR) weighting, adjust the untreated group to resemble the treated group by weighting or dropping members of the untreated group and leaving the members of the treated group untouched (i.e., given weights of 1). Some methods, such as pair matching without replacement, require the treated group to be smaller than the untreated group.

Pair matching works by assigning to each treated unit a matched untreated unit that is similar with respect to a distance metric, e.g., the difference between units' propensity scores. SMR weighting assigns to each treated a weight of 1 and to each untreated unit a weight of $\pi_i/(1-\pi_i)$, which makes the weighted distribution of covariates in the control group resemble that of the treated group. Stratification methods, such as full matching (15), propensity score subclassification (28), and fine stratification (16), involve assigning units into strata based on the covariate or propensity score values; a new "stratum propensity score" can be specified as the proportion of treated units in each stratum, and the formula for SMR weights can be applied to estimate weights that target the ATT using these methods (11).

It is critical that no treated units are dropped from the sample or given variable weights when estimating the ATT; doing so, such as by imposing a caliper or common support restriction, will change the estimand (29,30). Often, though, dropping treated units is required to achieve good balance, especially in the absence of good overlap (31). Many applications of pair matching in medical research use calipers (32,33), so it is important to interpret their results with caution and recognize that they may not be targeting the ATT.

Estimating the ATU

Methods that estimate the ATU are the same as those used for the ATT, replacing "treated" with "untreated" and vice-versa in the description above. That is, methods that estimate the ATU leave the untreated group untouched and use information from the treated group to simulate what would have happened had the untreated units been untreated; they do so by adjusting the treated group to resemble the untreated group. For SMR weighting for the ATU, untreated units receive a weight of 1 and treated units receive a weight of $(1 - \pi_i)/\pi_i$.

Estimating the ATE

Methods that estimate the ATE involve adjusting both the treated and untreated groups so that they each resemble the full sample. The most common method for estimating the ATE is inverse probability weighting, in which treated units receive a weight of $1/\pi_i$ and untreated units receive a weight of $1/(1 - \pi_i)$. Pair matching methods are typically inappropriate for the ATE because the resulting matched sample will resemble either the treated or untreated units (or neither if a caliper is used) but not the full sample, though there are variations that can target the ATE (34). The stratification methods mentioned previously can be naturally extended to target the ATE by applying the inverse probability weighting formula to the stratum propensity scores (11).

Estimating the ATO

Methods for estimating the ATO involve selecting or prioritizing treated and untreated patients who are very similar to each other and who have approximately equal chances of being treated or untreated; these units represent the equipoise population. Methods for the ATO typically have the potential for high precision and low bias because only the most similar units are retained or prioritized.

The most common methods for estimating the ATO include caliper matching (10,12) and overlap weighting (21,25). Caliper matching is a form of pair matching in which some treated units are left unmatched and dropped from the sample because no untreated units are within a specified maximum allowable distance (i.e., the caliper width) from them. Overlap weighting assigns to treated units a weight of $1 - \pi_i$ and to untreated units a weight of $\pi_i$. Coarsened exact matching (18) assigns units into strata based on coarsened versions of the covariates, which often involves dropping units that are in strata with no members of the other treatment group; when units from both treatment groups are dropped, the estimated effect corresponds to an ATO (35). Cardinality matching (36), which uses optimization to select the largest paired sample that satisfies user-specified balance constraints, can also target an ATO in cases of poor overlap or when the balance constraints are strict (20). Other weighting methods, such as matching weights (22) and weight trimming (29,37), also target the ATO.

When no specific estimand is of particular interest, methods used for estimating the ATO can be used because they often impart certain good statistical properties, such as precision (i.e., narrow confidence intervals), good covariate balance (i.e., treated and untreated units very similar to each other), and robustness to some forms of unmeasured confounding (38,39). What these methods have in common is that they prioritize patients far from the extremes and more similar to patients in the other group. These methods are therefore useful not just for specifically targeting the equipoise population but also for estimating a precise and robust treatment effect in *some* population, even if that population is not defined *a priori*. This is useful for treatment effect discovery, i.e., for deciding if at least some patients benefit from the treatment, which can be the impetus for a larger-scale study that does focus on larger or more specific patient populations (38).

### Cautions when interpreting estimates from other methods

Our focus has been on the estimates resulting from design-based methods like matching and weighting, but other methods, such as regression and instrumental variable analysis, are sometimes used to estimate treatment effects in observational studies. The estimates resulting from these methods generally do not correspond to any of the estimands discussed and instead target more nuanced estimands. In the Appendix, we summarize these methods and the implications for interpreting the estimands they target.

## DISCUSSION

Before analyzing an observational dataset, a researcher should consider which research question they want to ask and to which members of a target population the question refers. After making this decision, they should then choose the statistical method that corresponds to their chosen estimand. The effect estimate should be interpreted with respect to the patient population that corresponds to the estimand, which should ideally also be characterized based on the distribution of patient characteristics. Following these steps will ensure that studies examining the same phenomena are comparable and that their results can be correctly interpreted. Below, we give some practical considerations for how to choose a target estimand.

### Questions to ask when choosing an estimand

The first question to ask when choosing an estimand is of what kind of actual or implied policy the research question is trying to assess. If the question concerns a policy of withholding treatment from those who would currently receive it or preventing exposure for those likely to be exposed, this suggests the ATT is of interest. If the question concerns a policy of expanding treatment to those who would not currently receive it, the ATU may be of interest. If the question concerns a policy that would require all patients to be treated or to be untreated, the ATE may be of interest. If the question concerns a policy of prescribing treatment for patients under uncertainty, the ATO may be of interest. Deciding which type of question is the most useful one to ask depends on how the results of the study would be used by practitioners and other stakeholders and what utility the corresponding policy would have.

The second question to ask is which assumptions can be endorsed and justified by the researcher. Although conditional exchangeability, positivity, and SUTVA are required when using matching, weighting, or regression for estimating all estimands, the strictness of the assumptions differs depending on the estimand, as described in the Appendix. In particular, the ATE requires the strictest versions of these assumptions because it involves consideration of the unseen potential outcomes for the entire study population, whereas the other estimands only require unseen potential outcomes for narrower subsets of patients. There may be cases where the presence of unmeasured factors that affect treatment selection or the eligibility of some patients to be treated prohibit the use of certain estimands. These assumptions should be carefully considered using subject matter expertise and consultation with medical practitioners.

The third question to ask is whether the available data support the estimation of the desired estimand. Sometimes, the performance of statistical methods when applied to a dataset may not be adequate to robustly estimate the desired effect without strong modeling assumptions. This is especially common with smaller datasets and with datasets where the treated and untreated units differ substantially from each other. In these cases, there may be no matching or weighting specification that successfully equates the groups while retaining the desired estimand. Instead, it might be necessary to eschew the original estimand in favor of one that can be estimated with precision in these extreme cases. To figure out whether the statistical methods will be adequate for estimating the desired estimand, one should try several methods of matching and weighting without involving the outcome and assess the ability of each method to equate the groups on patient characteristics without discarding or down-weighting too many patients. If no method can provide a large enough and well-balanced matched or weighted sample, it may be necessary to change the estimand to one better supported by the data. Methods that target the ATO tend to perform better than methods targeting other estimands, whereas methods targeting the ATE tend to encounter problems more frequently because they require good overlap across the whole range of patients.

## Conditional Estimands and Subgroup Analysis

We have focused on estimands whose target populations depend on the treatment status of the patients under study; however, it is often valuable to know the treatment effect for specific patient populations defined by patient characteristics. Conditional estimands are estimands corresponding to specific subgroups of patients based on their baseline characteristics. One can consider any of the previously described estimands within a particular subgroup. For example, the ATT of a drug in a subgroup of male patients over 60 with hypertension is the effect of the drug for patients taking the drug and having these characteristics, while the ATE of the drug in that subgroup would be the effect of the drug for all members of that subgroup. Subgroup estimands tend to provide more useful information than overall population estimands because they reflect how providers typically make recommendations, i.e., based on known patient characteristics. For example, a treatment might be helpful on average for high-baseline severity patients and harmful for low-baseline severity patients; an overall ATE would miss this nuance, but the subgroup ATEs stratified by baseline severity would provide information more useful for decision-making by providers. Subgroup analyses may require special care when using matching or weighting (40,41).

## Conclusion

Different research questions may implicitly refer to different patient populations that correspond to specific estimands, and it is important that researchers be clear about their target estimand because the choice of statistical method used to estimate a treatment depends on the estimand of interest. Each estimand answers a different substantive question: the ATT answers a question about the effect of withholding treatment or preventing exposure, the ATU about expanding treatment, the ATE about implementing a universal treatment policy, and the ATO about selecting treatment under uncertainty. Different statistical methods estimate different estimands or need to be adjusted to correctly target the estimand of choice. There may be cases where a research question does not imply a specific patient population, in which case methods targeting any population can be used, but these scenarios should be clearly articulated, and it should be understood that results from such an analysis may not generalize or replicate beyond the sample at hand.

We hope that this paper provides guidance for medical researchers seeking to estimate the effect of a procedure, drug, or exposure. By clearly articulating the target estimand of an analysis, researchers can ensure their estimates correspond to a meaningful patient population and can choose the correct statistical method for their purposes.

## APPENDIX

### Outcome regression

As discussed in the main text, outcome regression, which involves fitting a model (e.g., a linear, logistic, or Poisson regression model) for the outcome as a function of the treatment, patient characteristics, and possibly the propensity score, is a commonly used alternative to matching and weighting. When the coefficient on the treatment in such a model is used as the estimate of the treatment effect, difficulties may arise with the interpretation of the effect with respect to the estimand.

The most typically used form of regression with a continuous outcome, the main effects model (also known as analysis of covariance, or ANCOVA), uses a model for the outcome with no interaction between the treatment and patient characteristics (42). This model implicitly assumes that the ICE is the same for all patients, an unrealistic assumption that, when false, makes the ANCOVA estimand correspond to an ambiguous weighted average of the ATT and ATU (43). This can be avoided by including interactions between the treatment and covariates in the model, but special procedures must be used to extract a treatment effect estimate from this more complicated model (42).

With logistic regression models for binary outcomes, the situation is worse due to the noncollapsibility of the odds ratio, an often-misunderstood issue in the interpretation of logistic regression results that is beyond the scope of this paper (44). The treatment coefficient in a main-effects logistic regression model corresponds to an ambiguous conditional estimand that is totally different from the one that matching and weighting methods target, making the results of logistic regression not comparable with results from matching and weighting.

It is possible, however, to use regression to validly target the ATT, ATU, or ATE with g-computation, also known as regression estimation or the parametric g-formula, which involves predicting the potential outcomes under each treatment level from a fitted outcome model and using them to estimate average treatment effects (42,45,46). G-computation can also be combined with propensity score-based methods to form "doubly-robust" estimators that are consistent if either the propensity score model or outcome regression is correctly specified (47,48).

### Instrumental Variable Analysis

Some medical studies use an instrumental variable, a pre-treatment variable that affects treatment assignment but is otherwise unassociated with the outcome, to adjust for confounding (49,50). Examples of instrumental variables in medical studies include the provider's preference for one method of care over another (51) and whether a patient was given a recommendation for particular health behavior (52). The estimand associated with instrumental variable analysis is the complier average treatment effect (sometimes "local" average treatment effect or LATE), which is the average of the ICEs for those who "comply with" (i.e., actually take) the treatment recommended to them (53). Typically, though, this is not the estimand of interest in medical studies because those who comply with respect to a specific instrument may be quite different from the eligible patient population (54). Estimates resulting from instrumental variable analysis are therefore not directly comparable to the results obtained from other methods and cannot be validly interpreted as the ATT, ATU, or ATE (24).

Assumptions for causal inference

The assumptions required to estimate average treatment effects using covariate adjustment-based methods like matching, weighting, and outcome regression include conditional exchangeability, positivity, and the stable unit treatment value assumption (SUTVA). These assumptions are briefly described in the main text in the section "Assumption for Causal Inference". The assumptions of conditional exchangeability and positivity differ depending on the target estimand; the assumptions are weaker for some estimands than for others. In addition to the substantive matters discussed in the main text, the assumptions a researcher is willing to endorse can be used to guide the choice of estimand. This section discusses how these assumptions vary with the estimand.

*Conditional Exchangeability*

The assumption of conditional exchangeability requires that a sufficient set of variables have been measured in the data and are included in the analysis. Formally, the assumption can be written as

$$Y^t \perp T|X$$

for $t = 0$ and $t = 1$ (8). Conditional exchangeability is a statement about treatment assignment: that the factors that influence the outcome under treatment are unrelated to treatment assignment conditional on $X$, and the factors that influence the outcome under no treatment are unrelated to treatment assignment conditional on $X$. This assumption, along with SUTVA, allows us to use the observed outcomes for one treatment group in place of the potential outcome for the other treatment group after adjusting for the covariates (e.g., using matching or weighting).

As written above, full conditional exchangeability is required for estimating the ATE. However, for the other estimands, the requirements are somewhat weaker. For the ATT, since the potential outcome under treatment is observed for the population of interest (the treatment group), the conditional exchangeability assumption involves only $Y^0$ (55):

$$Y^0 \perp T|X$$

With this assumption, only the factors that influence the outcome under no treatment must be unrelated to treatment assignment conditional on $X$. For example, if a surgeon's experience with a new surgical procedure affects the probability of a patient receiving the new procedure vs. standard surgery and would affect the patient's outcome under the new procedure but not under the standard surgery, the ATT can still be estimated even if the surgeon's experience with the new procedure is not measured, while other estimands could not be. Analogously, for the ATU, the conditional exchangeability assumption is

$$Y^1 \perp T|X$$

so that only the factors that influence the outcome under treatment must be unrelated to treatment assignment conditional on $X$. For the ATO, the conditional exchangeability assumption is

$$Y^t \perp T|X$$

for $t = 0$ and $t = 1$, but only within the equipoise population. The means unmeasured confounding can be present at the extremes, i.e., by factors that nearly guarantee either being treated or not being treated; for example, Stürmer et al. (39) identify that patient frailty, typically unmeasured, can cause patients to be essentially guaranteed treatment (i.e., as a last resort) or essentially guaranteed the absence of treatment (i.e., because the patient is too frail to receive it). Methods that target the ATO are not affected by the presence of unmeasured confounding present only in patients outside clinical equipoise, whereas targeting other estimands may yield biased estimates in these cases.

*Positivity*

The assumption of positivity requires that all patients are eligible to receive either treatment (8); it can be written as

$$0 < P(T = 1|X = x) < 1$$

for patients with all observed characteristic profiles $x$. The individual causal effect for any patients ineligible to be either treated or untreated cannot be defined because only one potential outcome is realizable. An estimand that involves both potential outcomes for those patients is essentially undefined and may not be able to be estimated. A lack of positivity manifests as treated and untreated populations that differ fundamentally from each other with no overlap on certain characteristics, making it impossible for matching or weighting to produce comparable groups.

The ATE requires the form of positivity written above, but other estimands require weaker forms and so may be more useful in cases where not all patients can receive one treatment or the other. For the ATT, the positivity assumption is

$$P(T = 1|X = x) < 1$$

for patients with all observed characteristic profiles $x$, which requires that all patients have the potential not to be treated. However, some patients can be ineligible to be treated (in which case they will likely not contribute much to the effect estimation). For the ATU, the positivity assumption is

$$P(T = 1|X = x) > 0$$

for patients with all observed characteristic profiles $x$, which requires that all patients are eligible to be treated, but some patients can have the potential not to be treated. If patients with high baseline disease severity were present among both treated and untreated groups but patients with low baseline severity were only present in the untreated group, the ATT could still be estimated (assuming other assumptions are met), but neither the ATE nor ATU could be (Figure A1).

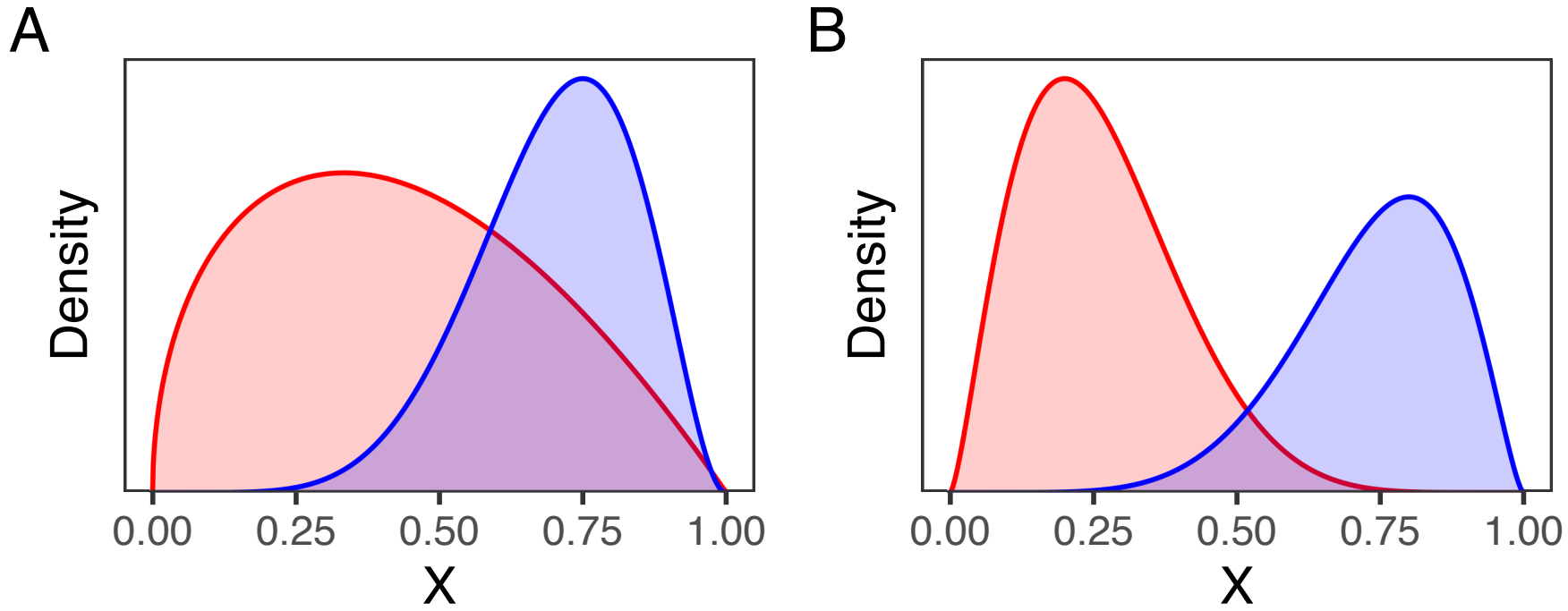

Figure A1. Plots of patient distributions as a function of a prognostic variable $X$ (e.g., baseline severity) that causes selection into treatment (blue) or no treatment (red). A) Patients with high values of $X$ are present in both the treated and untreated groups, but patients with low values of $X$ are only present in the untreated group, indicating a lack of overlap. The ATT can be estimated, but the ATE and ATU cannot be. B) Patients with high values of $X$ are present only in the treated group, and patients with low values of $X$ are present only in the untreated group, indicating a lack of overlap. Patients with moderate values of $X$ are present in both groups, so the ATO can be estimated, but the ATE, ATT, and ATU cannot be. The equipoise population would be defined as those with moderate values of $X$.

For the ATO, positivity requires that

$$0 < P(T = 1|X = x) < 1$$

for patients with at least some profiles $x$ (i.e., rather than all patients). The equipoise population is defined as patients with approximately equal probabilities of being treated or not treated, implying that they are both eligible to be treated and eligible not to be treated. For example, if there were some types of patients who always were treated and some types that were never treated, it may not be possible to estimate the ATE, ATT, or ATU, but as long as there was some area of overlap, with at least *some* patients both eligible to be treated and eligible not to be treated, the ATO could be estimated (Figure A1). Note that depending on the extent of overlap the ATO population may look quite similar to the ATE population, the ATT population, or the ATU population, or it may be a quite restricted subsample in cases with limited overlap.

Multi-category treatments

We have only considered binary treatments so far, but scenarios exist in which the treatment can take on one of multiple values, such as when considering two alternative treatments against a placebo or examining the effect of treatment dose (56). Estimands for multi-category treatments are interpreted similarly to how they are for binary treatments with a few exceptions (57). The ATE and ATO are each defined as the contrast between the expected potential outcomes under a pair of treatments for the full population or for an equipoise group, respectively.

For the ATT or ATU, one must consider which one of the treatment levels is to be the "focal" treatment level, i.e., the level that will serve as the target population for the estimand. For example, for a three-category treatment in which the levels are control, a low dose of a medication, and a high dose, the ATUs would correspond to the effect of low dose vs. control for those who would receive control and the effect of high dose vs. control for those who would receive control. The effect of low dose vs. high dose for those who would receive control is also a valid estimand and is equal to the contrast between the previous two ATUs. These estimands would be written as follows:

$$ATU_{l,c|c} = E[Y^l|T = c] - E[Y^c|T = c]$$

$$ATU_{h,c|c} = E[Y^h|T = c] - E[Y^c|T = c]$$

$$ATU_{l,h|c} = E[Y^l|T = c] - E[Y^h|T = c] = ATU_{l,c|c} - ATU_{h,c|c}$$

where $c$, $l$, and $h$ refer to control, low dose, and high dose, respectively. The ATTs for the low dose and high dose (i.e., the effect estimated for those who would receive the low dose and the effects estimated for those who would receive the high dose) are not comparable to each other because they reference different target populations.

Propensity score weighting and certain matching methods can be used to estimate treatment effects for multi-category treatments; see Lopez and Gutman (57) for a review. It is important to note that several of the described matching methods can discard units from all treatment groups, yielding an ATO rather than an ATT, ATU, or ATE. Ensuring the focal group remains intact is essential to being able to validly interpret an estimate as the average treatment effect in the focal group, though, as in the case of a binary treatment, this can yield bias in the absence of good overlap (57). The ATO can be estimated with weighting using a multi-category extension to overlap weights (58) or trimming (59).